\documentstyle[12pt]{article}

\newcommand{\sect}[1]{\setcounter{equation}{0}\section{#1}}

\newcommand{\be}{\begin{equation}}
\newcommand{\ee}{\end{equation}}

\newtheorem{teo} {Theorem} [section]
\newtheorem{defi}[teo] {Definition} 
\def\bkC{{\rm \kern.24em 
\vrule width.05em 
\vrule height1.4ex 
depth-.05ex 
\kern-.26em C}}

\begin{document}

\setlength{\oddsidemargin}{0cm} \setlength{\baselineskip}{7mm}

\begin{normalsize}\begin{flushright}
\end{flushright}\end{normalsize}

\begin{center}
  
\vspace{25pt}
  
{\Large \bf Simplicial minisuperspace models in the presence of a
scalar field }

\vspace{40pt}
  
{\sl Crist\'{o}v\~{a}o Correia da Silva}
$^{}$\footnote{e-mail address :
clbc2@damtp.cam.ac.uk}
and {\sl Ruth M. Williams}
$^{}$\footnote{e-mail address : rmw7@damtp.cam.ac.uk}
\\

\vspace{20pt}

DAMTP, Silver Street, Cambridge, \\
CB3 9EW, England \\

\end{center}

\vspace{40pt}

\begin{center} {\bf ABSTRACT } \end{center}
\vspace{12pt}
\noindent

We generalize simplicial minisuperspace models associated with
 restricting the topology
of the universe to be that of a cone over a closed connected combinatorial
$3-$manifold by considering the presence of a massive scalar
field. By restricting all the interior  edge lengths and all the
 boundary edge lengths to be
 equivalent and the scalar field to be homogenous on the $3-$space, we
 obtain a family of two dimensional models that include some of the most relevant
 triangulations of the  spatial universe. After studying the analytic
 properties of the action in the space of complex  edge lengths we
 determine its classical extrema.
 We then obtain   steepest
 descents contours of constant imaginary action passing through
 Lorentzian  classical geometries    yielding a convergent wavefunction
 of the universe, dominated by the contributions coming from these extrema. By considering these contours we justify   semiclassical approximations based on
 those classical solutions,  clearly  predicting classical spacetime in the
 late universe.  These wavefunctions are then evaluated numerically.
 For all of the models examined  we find wavefunctions predicting
 Lorentzian oscillatory behaviour in the late universe. 

\vspace{24pt}

\vfill

\newpage

\sect{Introduction}

The quantization of gravity, (QG), is perhaps the most important problem in
theoretical physics today. Among the several avenues proposed to
achieve that goal, one of the most productive has been the sum over
histories formulation. In such a formulation an amplitude for a
certain state of the universe is constructed by summing over a certain
class of physically distinct histories that satisfy appropriate
boundary  conditions, weighted by their respective  action. Problems
with the convergence of the path integral for gravity make it
convenient to use the Euclidean version of the sum over histories.

In the (Euclideanized) classical theory of gravity, i.e., general
relativity, (GR), a classical history is a Riemannian manifold, that
is, a smooth manifold $(M^{n},A)$ with smooth structure, (atlas), $A$,
endowed with a Riemannian metric $g$. Topological manifolds $M^{n}$,
with their local homoeomorphisms to $R^{n}$, are the mathematical
implementation of GR's Equivalence Principle. The smooth and metric
structures are essential for the definition of the most basic physical
concepts like distance, curvature, field derivatives, etc.

When building a quantum version of GR there are several reasons why we
should consider the generalisation of this concept of history. Any sum
over histories can only be implemented once we specify what are the
elements of the set of (physically distinct) histories in QG we have
chosen to consider. Such choice is constrained by several criteria
that any reasonable history in QG must meet. The most fundamental are:

\begin{itemize}

\item  It should be based on a finite dimensional topological space
$X^{n}$ endowed with a smooth structure $A$.

\item   The above mentioned space must be metrizable.

\item   The set of physically distinct histories $\{X^{n},A,g_{\mu
\nu}\}$ considered in the sum must be algorithmically decidable, AD, and
classifiable, AC.

\item   The action functional used should coincide with the usual
 Einstein action for manifold-based histories.
 
\end{itemize}

Given such a set of histories it is then possible to write a
probability amplitude for topology change between a set of boundary
$(n-1)-$manifolds $\partial X^{n-1}$ in the same cobordism class, as 

\be
\label{1}
G[\partial X^{n-1},h]=\sum _{(X^{n},A)} \int Dg_{\mu
\nu}e^{-I[X^{n},A,g_{\mu \nu}]}
\ee

where the sum is over all physically distinct histories, $\{X^{n},A,g_{\mu
\nu}\}$, that have the appropriate boundaries $(\partial X^{n-1})$,
and induce the desired $(n-1)$ metric, $h$, on those same
boundaries, and $I$ is the Euclidean action associated with each history.

Any successful theory of QG must have GR as its low energy limit, and
so its space of histories must contain the space of classical,
manifold-based histories. However, although necessary there are
several indications that they are not sufficient. To start with, the
set of $n-$dimensional smooth manifolds is not AC for $n\geq 4$, and
there is no known algorithm to decide whether or not a generic
topological space is a manifold in $4$ dimensions. It was
proven there is no way to do so for higher dimensions.

Furthermore, there are cases in which formal descriptions of histories
based on the classical configurations of a theory, do not necessarily
correspond to the precise mathematical definition of the space of
histories needed in order to make Euclidean sums over histories both
well defined  and yielding the correct quantum mechanics, \cite{h0}. 
 On the other hand these histories cannot be based on very
pathological topological spaces, because it would be impossible to
concretely define the action associated with those generalized
topological spaces.
This requires the definition of concepts like distance, volume and
scalar curvature. In order to define distance, the topological spaces
must be metrizable, and it can be shown that  a notion of curvature
can be introduced in any metrizable space where such notion of
distance exists.

However, such spaces must also have uniform dimension, otherwise it
would not be clear how to weight  their contributions  in a
sum over histories, since the form and properties of the action depend
on the dimension of the space. As pointed out by Scheleich and Witt in
\cite{witt1}, these restrictions are still not enough to eliminate all
unsatisfactory topological spaces, and they go on to discuss in some
detail what other restrictions are necessary. 

 A good candidate for this generalized space of histories as been
suggested in \cite{witt1}. They propose generalizing from  manifold-based
histories to conifold-based histories. Briefly a conifold is a
topological space that is like a manifold everywhere except in a
discrete, countable set of  points $S$, (called singular points), that
do not have any neighbourhood homeomorphic to an open ball of $R^{n}$,
but to a cone over some closed connected $(n-1)-$manifold (other than
$S^{n-1}$, of course).

In one and two dimensions, manifolds and conifolds coincide. But for
higher dimensions the set of conifolds more than contains the set of
manifolds. The fact that the set of non-manifold like points, S, in
any conifold is discrete and countable allows the direct
generalization of all basic geometrical concepts, (such as geodesics,
curvature, etc.) from manifolds to conifolds by continuation.

The problem now is how to concretely implement such a sum. To do it we
must not only define the space of histories we will be considering but
also obtain a finite representation for them. A simplicial formulation
of QG in terms of Regge calculus aims to provide one such
representation.
 Following \cite{witt1} by defining combinatorial manifolds and
conifolds
in the appropriate way, we can be sure that in less than seven
dimensions there is a one to one relationship between smooth
manifolds/conifolds and their combinatorial counterparts.

 These combinatorial counterparts are thus a  finite representation of
the smooth topological spaces they are  associated with. Furthermore
 this one to one relationship allows us to
substitute the sum $(\ref{1})$ for a sum over those simplicial complexes,
which is easier to concretely implement.

However, the computation of the full sum is not at present viable, and
so we restrict ourselves to a simplicial minisuperspace approximation
for the wavefunction of the universe for some simple but fairly
general models. Such calculations were initiated in \cite{h1},
\cite{h2} and extended
in \cite{birm} and \cite{fur}. We  aim  to further extend these
models and verify if
they yield  a wavefunction predicting classical spacetime.

Our simplicial minisuperspace approximation consists in restricting the
topology of the $4-D$ universe to be that of a simplicial cone over a
$3-D$ closed connected combinatorial manifold, which is to be seen as 
a triangulation of the spatial $3-D$ universe. Furthermore,  its
geometry must  be
such that there is a single type of boundary edge ,with square length
$s_{b}$, and internal edge with squared length, $s_{i}$. We shall also
 consider that there is a massive scalar field 
present, but that it takes the same value in all the vertices of the
spatial $3-D$ universe, which is analogous to the requirement of the
scalar field to be homogeneous $\phi =\phi (t)$ in the usual continuum
minisuperspace models, \cite{hawk}.

As pointed out in \cite{h2}, we shall see that the only way of obtaining a
wavefunction predicting a late universe exhibiting classical spacetime
like our own, (i.e. a wavefunction with Lorentzian oscillatory
behaviour), 
 is to consider contours of integration over complex valued interior edge
lengths, $s_{i}$. This forces us to study the minisuperspace action as
a function of complex variable, which is rather involved because of
the multivaluedness of the action.

 We then try to find convergent integration contours yielding appropriate
 wavefunctions, i.e., that predict Lorentzian classical spacetime
 in the late universe, by allowing a semiclassical approximation based on
 Lorentzian classical  solutions to be a good approximation of the
 full wavefunction when the universe is large.

\sect{Simplicial Quantum Gravity}

\subsection{Combinatorial Manifolds and Conifolds}

Before we introduce our simplicial minisuperspace model it is convenient
 that we review some basic definitions of simplicial geometry.

\begin{defi}

 A simplicial complex $(K,\mid K\mid )$ is a topological space $\mid K\mid$ and a
 collection of simplices $K$,such that

\begin{itemize}

   \item $\mid K\mid$ is a closed subset of some finite dimensional
   Euclidean space.

   \item  If $\sigma $ is a face of a simplex in $K$, then $\sigma $
   is also contained in $K$.

   \item  If $\sigma _{a}$ and $\sigma _{b}$ are simplices in $K$, then  
   $\sigma _{a} \cap \sigma _{b}$ is a face of both  $\sigma _{a}$
   and $\sigma _{b}$.

   \item The topological space $\mid K\mid $ is the union of all
   simplices in $K$.
\end{itemize}     
\end{defi}

\begin{defi}
 
A combinatorial $n-$manifold ${\cal M}^{n}$, is an $n-$dimensional simplicial complex such  that  
       
\begin{itemize}

     \item It is pure.
     \item It is non branching.
     \item Any two $n-$simplices can be connected by a sequence of $n-$simplices, each intersecting along some $(n-1)-$simplex.
     \item  The link of every vertex is a combinatorial $(n-1)-$sphere.

\end{itemize}
\end{defi}

Note that there are simplicial complexes that are homeomorphic to topological manifolds but are not combinatorial manifolds. The definition of combinatorial manifold carries more structure than simply the topology. A similar definition can be made for combinatorial conifolds ${{\cal C}}^{n}$, by simply replacing the  previous last condition by 

\begin{itemize}  

\item The link of every vertex of ${{\cal C}}^{n}$ is a closed connected combinatorial $(n-1)-$manifold.

\end{itemize}

Thus, like in the continuum framework, a general combinatorial conifold is a combinatorial manifold everywhere except possibly at a countable set of vertices,

Combinatorial manifolds and conifolds are  in a similar situation
to smooth manifolds and conifolds, in the sense that they both have
additional structure relative to their more general topological
counterparts. Such structure  allows integration and differentiation to be well
defined as essential for any physical applications. In order to see the
connection between the smooth and combinatorial structures we need to
introduce one more definition:

\begin{defi}

A combinatorial triangulation of a manifold, $M^{n}$,
consists of a combinatorial manifold  ${{\cal M}}^{n}$, and an
homeomorphism 
$t:\mid {{{\cal M}}^{n}}\mid  \rightarrow  M^{n}$.
\end{defi}
An analogous definition holds for conifolds.

Following \cite{witt1} it can be
shown that every smooth manifold and conifold  admit combinatorial
triangulations. On the other hand, it can be also shown that every
combinatorial manifold admits  a piecewise-linear, (PL), structure, i.e. a
PL-atlas, $\{(V_{a},\phi _{a})\}_{a\in A}$, such that the mappings

\be
\phi_{b}\phi_{a}^{-1}:\phi_{a}(V_{a}\cap V_{b})\rightarrow \phi_{b}(V_{a}\cap V_{b})
\ee
are PL mappings between subsets of $R^{n}_{+}$.
Finally it is known   that in less than seven dimensions every manifold
with a 
PL-structure has an unique smoothing. A similar reasoning can be
applied to   conifolds, and
so we are able to conclude that

\vspace{20pt}

   In less than seven dimensions,  smooth manifolds/conifolds,
   $(M^{n}/C^{n})$, uniquely correspond to  combinatorial manifolds/conifolds, $({{\cal M}}^{n}/{{\cal C}}^{n})$. 

\vspace{20pt}  

Obviously  the topological space underlying each smooth  manifold/conifold may have several inequivalent
triangulations, what this result says is that when we specify a smooth
structure  on that space it  will correspond to a unique 
combinatorial triangulation, i.e., a triangulation based on a
combinatorial manifold/conifold, and not just any simplicial complex.
This unique correspondence implies that in less than seven dimensions
the information about the smooth structure of the space is carried by
its combinatorial triangulation.
So we see that the topological part of the sum over histories in
$(\ref{1})$, can be recast in terms of simplicial representatives of
the ``continuum'' spaces. However in order to fully translate an
heuristic expression such as $(\ref{1})$, into a concrete sum over simplicial histories we still need to associate a metric and an action to each simplicial complex to be considered. Note that up to now we have not specified any kind of metric information associated with simplicial complexes. Once we have fixed the topology of the underlying simplicial complex, the most convenient way to attach metric information to it, is to use Regge calculus.

\subsection{General Regge Formalism}

A convenient way of defining an $n-$simplex is to specify the
coordinates of its $(n+1)$ vertices, $\sigma =[0,1,2,...,n]$. By
specifying the squared values of the lengths of the edges $[i,j]$,
$s_{ij}$, we fix the simplicial metric on the simplex.

\be
\label{simpmet}
g_{ij}(s_{k})=\frac{s_{0i}+s_{0j}-s_{ij}}{2}
\ee
where $i,j=1,2,..n$.

So if we triangulate a smooth manifold $M$ endowed with a metric $g_{\mu \nu}$ by a homeomorphic simplicial manifold 
${\cal{M}}$, the metric information is transferred to the simplicial metric of that simplicial complex

\be 
g_{\mu \nu}(x) \longrightarrow   g_{ij}(\{s_{k}\})=\frac{s_{0i}+s_{0j}-s_{ij}}{2}
\ee

In the continuum framework the sum over metrics is implemented through a functional integral over the metric components $\{g_{\mu \nu}(x)\}$. In the simplicial framework the metric degrees of freedom are the squared edge lengths, and so the  functional integral is replaced by a simple multiple integral over the values of the edge lengths. But not all edge lengths have equal standing. Only the ones associated with the interior of the simplicial complex get to be integrated over. The boundary edge lengths remain after the sum over metrics and become the arguments of the wavefunction of the universe.

\be
\int Dg_{\mu \nu }(x) \longrightarrow \int D\{s_{i}\}=\prod \int d\mu (s_{i}) 
\ee

In the simplicial framework the fact that the geometry of the complexes is completely fixed by the specification of the squared values of all edge lengths, means that all geometrical quantities, such as volumes and curvatures, can be expressed completely in terms of those edge lengths. Consequently the Regge action (the simplicial analogue of the Einstein action for GR) associated with a complex of known topology can be expressed exclusively in terms of those edge lengths.

The Euclideanized Einstein action for a smooth $4-$manifold $M$ with boundary $\partial M$, and endowed with a $4-$metric, $g_{\mu \nu }$, and a scalar field $\Phi $ with mass $m$, is

\begin{eqnarray*}
I[M,g_{\mu \nu},\phi ]&=&-\int _{M}d^{4}x\sqrt{g}\frac{(R-2\Lambda
                       )}{16\pi G}-\int _{\partial
                       M}d^{3}x\sqrt{h}\frac{K}{8\pi G}+ \\
                       &+& \mbox{} \frac{1}{2}\int  _{M}d^{4}x\sqrt{g}(\partial _{\mu}\phi \partial ^{\mu} \phi +m^{2} \phi ^{2})   
\end{eqnarray*}
where $K$ is the extrinsic curvature.

Its simplicial analogue will be the Regge action for a combinatorial
$4-$manifold, ${\cal {M}}$, with squared edge lengths $\{s_{k}\}$, and
with a scalar field taking values $\{\phi _{v}\}$ for each vertex $v$
of ${\cal {M}}$, \cite{ruth3}:

\begin{eqnarray*}
 I[{\cal {M}},\{s_{k}\},\{\phi _{v}\}]&=&\frac{-2}{16\pi G}\sum _{\sigma _{2}^{i}} V_{2}(\sigma _{2}^{i})\theta (\sigma _{2}^{i}) +  {\frac{2\Lambda }{16\pi G}} \sum _{\sigma _{4}}V_{4}(\sigma _{4}) \\
                                      &-&{\frac{2}{16\pi G}}\sum _{\sigma _{2}^{b}} V_{2}(\sigma _{2}^{b})\psi (\sigma _{2}^{b})+\frac{1}{2}\sum _{\sigma _{1}=[ij]} \tilde{V}_{4}(\sigma _{1})\frac{(\phi _{i}-\phi _{j})^{2}}{s_{ij}}\\
                                      &+& \mbox{}\frac{1}{2}\sum _{j}\tilde{V}_{4}(j)m^{2}\phi _{j}^{2}   
\end{eqnarray*}
where:

\begin{itemize}

 \item $\sigma _{k}$ denotes a $k-$simplex belonging to the set $\Sigma _{k}$ of all  $k-$simplices in  ${\cal {M}}$.

\item $\theta(\sigma _{2}^{i})$, is the deficit angle associated with the interior $2-$simplex $\sigma _{2}^{i}=[ijk]$

\be
  \theta(\sigma _{2}^{i})=2\pi -\sum _{\sigma _{4}\in St(\sigma _{2}^{i})}\theta _{d}(\sigma _{2}^{i},\sigma _{4})   
\ee

and $\theta _{d}(\sigma _{2}^{i},\sigma _{4})$ is the dihedral angle between the $3-$simplices $\sigma _{3}=[ijkl]$ and $\sigma ^{'}_{3}=[ijkm]$, of $\sigma _{4}=[ijklm]$ that intersect at  $\sigma _{2}^{i}$. Its full expression is given by \cite{birm}.

\item   $\psi (\sigma _{2}^{b})$ is the deficit angle associated with the boundary $2-$simplex $\sigma _{2}^{b}$:

\be
  \psi (\sigma _{2}^{b})=\pi -\sum _{\sigma _{4}\in St(\sigma _{2}^{b})}\theta _{d}(\sigma _{2}^{b},\sigma _{4})    
\ee

\item $V_{k}(\sigma _{k})$ for $k=2,3,4$ is the $k-$volume associated with the $k-$simplex, $\sigma _{k}$, and once again their explicit expressions in terms of the squared edge lengths are given by  \cite{birm}.

\item  $\tilde{V}_{4}(\sigma _{1})$, is the $4-$volume in the simplicial complex ${\cal{M}}$, associated with the edge $\sigma _{1}$, i.e., the volume of the space occupied by all points of ${\cal{M}}$ that are closer to $\sigma _{1}$ than to any other edge of  ${\cal{M}}$. The same holds for $\tilde V_{4}(j)$ where $j$ represents all vertices of ${\cal{M}}$.

\end{itemize}

It is easy to see that both $\tilde{V}_{4}(\sigma _{1})$ and $\tilde
V_{4}(j)$, can be expressed exclusively in terms of the edge lengths
$\{s_{k}\}$. All  these expressions remain valid if we consider smooth conifolds and their combinatorial counterparts.

So we see that any reasonable history in QG of the type $(X^{4},A,g_{\mu \nu},\Phi )$, where $X^{4}$ represents either a topological manifold or conifold endowed with a smooth structure $A$, metric $g_{\mu \nu}$, and in the presence of matter fields represented by $\Phi$, has an {\it{unique}} simplicial analogue, $({\cal{X}}^{4},\{s_{k}\},\{\Phi _{j}\})$. This allows us to recast the heuristic expression $(\ref{1})$in terms of this finite representation as:

\be
\label{2}
\Psi [\partial {\cal{X}},\{s_{b}\},\{\phi _{b}\}]=\sum _{{\cal{X}}^{4}}\int D\{s_{i}\}D\{\phi _{i}\}e^{-I[{\cal{X}}^{4},\{s_{i}\},\{s_{b}\},\{\phi _{i}\},\{\phi _{b}\}]}  
\ee
where

\begin{itemize}

\item $ \{s_{i}\}$ are the squared lengths of the interior edges 

\item  $\{s_{b}\}$ are the squared lengths of the boundary  edges 

\item  $\{\phi_{i}\}$ are the values of the field at the interior vertices 

\item  $\{\phi_{b}\}$ are the values of the field at the boundary vertices

\end{itemize}

Although the functional integral over metrics has been written explicitly
 in terms of the edge lengths, this expression is still
heuristic because we still need to specify  the list of suitable
simplicial complexes ${\cal{X}}^{4}$ we intend to sum over,  the
measure,  and the integration contour to be used. One way to avoid the problems in defining one such
list, namely the problems of algorithmic decidability and
classifiability, (that have been discussed extensively in
\cite{witt1}) is to evaluate the sum approximately by singling out a 
subfamily of simplicial histories described by only a few parameters
and carrying out the sum over these histories alone.

 An example of this is to adopt a simplicial minisuperspace approximation. We now describe in some detail the minisuperspace model we shall consider.

\sect{Simplicial Minisuperspace}

We shall reduce our attention to a significant  subfamily of simplicial histories characterized by the following restrictions: 

\subsection{Topological Restrictions}

   We shall consider that the universe is well approximated as a simplicial cone ${\cal {C}}^{4}=a*{\cal{M}}^{3}$ over a $3-$dimensional closed combinatorial manifold ${\cal {M}}^{3}$, that is taken to triangulate the spatial $3-$dimensional universe. Models of this kind have already been considered in \cite{birm}.   

The most distinctive feature of such  a model is the simplifications introduced by the existence of only one interior vertex, the apex of the cone.

 The consequences of this restriction are

\begin{itemize}

\item  Note that the simplicial $4-$complex, ${\cal {C}}^{4}$, in
 gene\-ral will not be a com\-bina\-to\-rial $4-$mani\-fold but a com\-bina\-to\-rial
 $4-$conifold, because all the vertices in ${\cal {C}}^{4}$ 
are manifold-like points, except the apex, $a$, whose link
 is $L(a)={\cal{M}}^{3}$, and in general ${\cal{M}}^{3}$ 
 will not be a combinatorial $3-$sphere.

\item   The fact that ${\cal{M}}^{3}$ is closed means that the only boundary of ${\cal {C}}^{4}$ will be the ${\cal{M}}^{3}$ itself:

\be
\partial {\cal {C}}^{4}={\cal{M}}^{3}
\ee

So the combinatorial $3-$manifold ${\cal{M}}^{3}$ is to be seen as a triangulation of the spatial $3-D$ universe.

\item The cone structure of the ${\cal {C}}^{4}$ reflected in the fact that there is only one interior vertex (the apex $a$) means that all vertices in ${\cal {M}}^{3}$ will be boundary vertices of ${\cal {C}}^{4}$. 
So if $N_{p}({\cal {K}}^{n})$ is the number of $p-$simplices in the complex 
${\cal {K}}^{n}$ we see that since ${\cal {C}}^{4}=a*{\cal{M}}^{3}$, then

\be
 N_{0}({\cal {C}}^{4})=N_{0}({\cal {M}}^{3})+1
\ee

And in general there will be two kinds of $p-$simplices $(p=1,2,3)$,
in ${\cal {C}}^{4}$, the ones that exist originally in
${\cal{M}}^{3}$,  which will be the boundary $p-$simplices 
of ${\cal {C}}^{4}$, and the $p-$simplices generated as cones over the $(p-1)-$simplices of ${\cal{M}}^{3}$
with apex $a$. These will be the interior $p-$simplices of ${\cal {C}}^{4}$. So in general

 $$\sigma _{p }^{bound} ({\cal {C}}^{4})=\sigma _{p }({\cal{M}}^{3})$$
 $$\sigma ^{inter}_{p} ({\cal {C}}^{4})=a*\sigma _{p-1 }({\cal{M}}^{3})$$

and  

$$          N_{0}({\cal{C}}^{4})=N_{0}({\cal{M}}^{3})+1  $$
$$          N_{1}({\cal{C}}^{4})=N_{1}({\cal{M}}^{3})+N_{0}({\cal {M}}^{3})$$
$$          N_{2}({\cal{C}}^{4})=N_{2}({\cal{M}}^{3})+N_{1}({\cal {M}}^{3})$$
$$          N_{3}({\cal{C}}^{4})=N_{3}({\cal{M}}^{3})+N_{2}({\cal {M}}^{3})$$
$$          N_{4}({\cal{C}}^{4})=N_{3}({\cal {M}}^{3})$$

\item The Euler characteristic of a simplicial complex ${\cal{K}}^{n}$ is

$$ \chi ({\cal{K}}^{n})=N_{0}({\cal{K}}^{n})-N_{1}({\cal{K}}^{n})+N_{2}({\cal{K}}^{n})-N_{3}({\cal{K}}^{n})...\pm N_{n}({\cal{K}}^{n}) $$
with $+$ if $n$ is even and $-$ if $n$ is odd. On the other hand, since ${\cal{M}}^{3}$ is a closed combinatorial manifold, then its Euler characteristic must vanish, and so we have 

\be
N_{0}({\cal{M}}^{3})-N_{1}({\cal{M}}^{3})+N_{2}({\cal{M}}^{3})-N_{3}({\cal{M}}^{3})=0    
\ee

\item Since ${\cal{M}}^{3}$ is a closed, pure, non branching complex then all its  $2-$simplices must belong to exactly two $3-$simplices of   ${\cal{M}}^{3}$ 

\be
N_{2}({\cal{M}}^{3})=2N_{3}({\cal{M}}^{3})   
\ee

\end{itemize}

In the table below we list the values of $N_{0}$, and $N_{3}$ for some of the most
relevant closed connected combinatorial $3-$manifolds ${\cal{M}}^{3}$

\vspace{55pt}

\begin{tabular}{|c|c|c|c|}  \hline 
\hspace{.1in}${\cal{M}}^{3}$\hspace{.4in}&\hspace{.4in}$N_{0}$\hspace{.4in}
  &\hspace{.4in}$N_{3}$\hspace{.4in}
  &\hspace{.4in}$S_{crit}^{m=0}$\hspace{.4in} \\ \hline
\hspace{.4in}$\alpha _{4}$\hspace{.4in}   & \hspace{.4in}5 \hspace{.4in}      & \hspace{.4in}  5\hspace{.4in}  &\hspace{.4in}$29.31$\hspace{.4in}      \\ \hline
\hspace{.4in}${\cal S}^{2}\times {\cal S}^{1}$\hspace{.4in}   & \hspace{.4in}10\hspace{.4in}      & \hspace{.4in} 30\hspace{.4in} &\hspace{.4in}$5.61$\hspace{.4in}        \\ \hline
\hspace{.4in}${\cal L}(2,1)$\hspace{.4in} & \hspace{.4in}11\hspace{.4in}      & \hspace{.4in} 30\hspace{.4in} &\hspace{.4in}$3.54$\hspace{.4in}        \\ \hline
\hspace{.4in}${\cal L}(5,1)$\hspace{.4in} & \hspace{.4in}15\hspace{.4in}      & \hspace{.4in} 89\hspace{.4in} &\hspace{.25in}$-0.84$\hspace{.4in}        \\ \hline
\hspace{.4in}${\cal T}^{3}$\hspace{.4in}  & \hspace{.4in}15\hspace{.4in}      & \hspace{.4in} 90\hspace{.4in}  &\hspace{.25in}$-0.31$\hspace{.4in}       \\ \hline

\end{tabular}

\vspace{45pt}
where 

\begin{itemize}

\item $\alpha _{4}$ is the simplest triangulation of $S^{3}$,
\cite{h2}.

\item ${\cal S}^{2}\times {\cal S}^{1}$ and ${\cal T}^{3}$ are
simple triangulations of the spaces $S^{2}*S^{1}$ and $T^{3}$
constructed in \cite{book}.

\item ${\cal L}(p,1)$ are  simple triangulations of their respective
 Lens spaces, $L(p,1)$ used in \cite{birm}.

\end{itemize}
 The meaning and relevance of the values in the last column will be explained later.
\subsection{Metric Restrictions}

Up to now we have  concentrated on the restrictions on the topology of the simplicial spacetimes that characterize our minisuperspace models. The restrictions on the metric degrees of freedom are as important.

By assuming a cone-like structure ${\cal {C}}^{4}=a*{\cal {M}}^{3}$, we see that all the interior edges of ${\cal {C}}^{4}$ are of the same type ,i.e., one of the vertices is a boundary vertex belonging to ${\cal {M}}^{3}$, and the other is the (single) interior vertex of ${\cal {C}}^{4}$, its apex.

If we label the interior vertex as $0$ and the other (bounda\-ry) ver\-ti\-ces of ${\cal {C}}^{4}$ as $1,2,...,\-N_{0}({\cal {M}}^{3})$. Then the cone-like structure of ${\cal {C}}^{4}$ leads to all interior edges being of the same form $[0,\alpha]$, with $\alpha =1,2,..,N_{0}({\cal {M}}^{3})$.

So it makes sense to introduce the restriction that all interior edges have equal lengths whose squared value is denoted $s_{i}=s_{0\alpha }$. A similar assumption is made with respect to the boundary edge lengths, i.e., we consider them all to be equal to a common value $s_{ij}=s_{b}$, with $i,j=1,2,..,N_{0}({\cal {M}}^{3})$.

We are well aware that these simplifications greatly reduce the scope
of the model, namely because they result in there being only one kind
of $4-$simplex in the complex. Thus any real classical solution will
necessarily be either purely Lorentzian or Euclidean. However, it is well
known that in the analogous continuum minisuperspace model, complex
classical solutions not only exist but  in the late Universe even 
dominate the path integral, \cite{hawk}. In order to begin to circumvent this
problem we are currently studying a similar model with two different
interior edge lengths.

\subsection{Scalar Field}

The simplifications assumed in respect to the edge lengths makes it natural to assume that the scalar field is spatially homogeneous. So we assume that the scalar field takes the same value $\phi _{b}$ for all boundary vertices of ${\cal {C}}^{4}$. The value at the interior vertex, $\phi _{i}$, is independent.

\subsection{Minisuperspace Wavefunction}

We can now concretely implement a simplicial minisuperspace approximation to the wavefunction of the universe of the type $(\ref{2})$, as

\be
\label{ppsi}
\Psi [{\cal{M}}^{3},s_{b},\phi _{b}]=\int ds_{i}d\phi _{i}e^{-I[a*{\cal{M}}^{3},s_{i},s_{b},\phi _{i},\phi _{b}]}   
\ee

The Regge action for this minisuperspace can now be calculated. For simplicity we introduce rescaled metric variables:

\be
 \xi =\frac{s_{i}}{s_{b}}
\ee
\be
 S=\frac{H^{2}s_{b}}{l^{2}}
\ee
where $H^{2}=l^{2}\Lambda /3$, and $l^{2}=16\pi G$ is the Planck length. We shall work in units where $c=\hbar =1$.

Then the volume of the $4-$simplices in  ${\cal {C}}^{4}$ is

\be
 V_{4}(\sigma _{4})=\frac{l^{4}}{24\sqrt{2}H^{4}}S^{2}\sqrt{\xi -3/8}
\ee

The volume of the $N_{1}({\cal{M}}^{3})$ internal $2-$simplices, $\sigma _{2}^{i}$ in ${\cal {C}}^{4}$ is

\be
V_{2}(\sigma _{2}^{i})=\frac{l^{2}}{2H^{2}}S\sqrt{\xi -1/4}  
\ee

The volume of the $2N_{3}({\cal{M}}^{3})$ boundary $2-$simplices, $\sigma _{2}^{b}$ in  ${\cal {C}}^{4}$ is

\be
V_{2}(\sigma _{2}^{b})=\frac{\sqrt{3}l^{2}}{4H^{2}}S 
\ee

The volumes of the internal and boundary $3-$simplices of ${\cal {C}}^{4}$ are, respectively
\be
  V_{3}(\sigma _{3}^{i})=\frac{l^{3}}{12H^{3}}S^{3/2}\sqrt{3\xi -1} 
\ee
\be
  V_{3}(\sigma _{3}^{b})=\frac{\sqrt{2}l^{3}}{12H^{3}}S^{3/2}   
\ee

and so the dihedral angle associated with each internal $2-$simplex is

\be
\theta (\sigma _{2}^{i})=\arccos{\frac{2\xi -1}{6\xi -2}};  
\ee

for the boundary $2-$simplices we have  

\be
\theta (\sigma _{2}^{b})=\arccos {\frac{1}{2\sqrt{6\xi -2}}}.  
\ee

With respect to the matter terms, the kinetic term vanishes when the
edges $\sigma _{2}$ are boundary edges. The only non vanishing
contribution comes from the internal edges $\sigma _{2}=[0j]$.

Computing the relevant volumes associated with the internal edges and
all the vertices it is possible to conclude that the  Regge action for
this simplicial minisuperspace is

\begin{eqnarray*}
I[\xi ,S,\phi _{i},\phi _{b}]&=&-\frac{S}{H^{2}}\biggl \{ (N_{3}\sqrt{3})\biggl[\pi -2\arccos {\frac{1}{2\sqrt{6\xi -2}}}\biggr] \\
                                   &+&N_{1}\sqrt{\xi -1/4}\biggl[2\pi -\frac{6N_{3}}{N_{1}}\arccos{\frac{2\xi -1}{6\xi -2}}\biggr]  \\
                                   &-& \biggl(\frac{N_{3}}{120\sqrt{2}}\biggr)\frac{\sqrt{\xi -3/8}}{\xi}(\phi _{i}l-\phi _{b}l)^{2}\biggl\} +\frac{S^{2}}{H^{2}}\biggl\{\biggl(\frac{N_{3}}{4\sqrt{2}}\biggr)\sqrt{\xi -3/8}  \\  
                                   &+& \mbox{} \frac{N_{3}}{240\sqrt{2}}\biggl(\frac{m^{2}l^{2}}{H^{2}}\biggr)\sqrt{\xi -3/8}(\phi _{i}^{2}l^{2}+4\phi _{b}^{2}l^{2})\biggr\}
\end{eqnarray*}
\begin{equation}
\label{actionI}
\end{equation}
where $N_{0},N_{1}$ and $N_{3}$ all refer to  ${\cal{M}}^{3}$.

We see that the dependence of the action on the topology of the
underlying ${\cal{M}}^{3}$ is contained in the parameters $N_{0}$ and
$N_{3}$, $(N_{1}=N_{0}+N_{3})$. The metric dependence is obviously contained in $\xi $ and $S$, and its dependence on the matter field in $\phi _{i}$ and $\phi _{b}$.

Note that the previous expressions are valid for a wide variety of simplicial geometries with very different spatial topologies (of ${\cal{M}}_{3}$). Different triangulations of the $3-D$ spatial universe have different values of $N_{0}$ and $N_{3}$ and consequently different actions. But as long as they are closed connected combinatorial manifolds the previous expressions hold and so the functional dependence of the action on $s_{i}$ and $s_{b}$ remains the same.

So given a certain underlying simplicial complex $a*{\cal{M}}^{3}$ (with fixed  $N_{0}$ and $N_{3}$ ) we see that in order to approximate the heuristic expression $(\ref{1} )$ by a fully computable expression 

\be
\Psi [{\cal{M}}^{3};s_{b};\phi _{b}]=\Psi [N_{0},N_{3};S;\phi _{b}]=\int _{C}D\xi D\phi _{i}e^{-I [N_{0},N_{3};S,\xi ;\phi _{b},\phi _{i}]}  
\ee

we only need to specify the integration contour $C$, and the measure of integration $D\xi D\phi _{i}$.

 In our simplified models  the result yielded by a contour C is not
 very sensitive to the choice of measure if we stick to the usual
 measures, i.e., polynomials of the squared edge lengths. In our case we take 

\be
 D\xi D\phi _{i}=\frac{ds_{i}}{2\pi il^{2}}d\phi _{i}=\frac{S}{2\pi iH^{2}}d\xi d\phi _{i}  
\ee

In the search for the correct integration contour for our simplicial minisuperspace model we start by reviewing some of the main results for the general continuum case.

Unfortunately,  in the case of closed cosmologies there is as yet no known explicit prescription for the integration contour. So usually one takes a pragmatic view, in which we look for contours that lead to the desired features of the wavefunction of the universe. Following \cite{hh}, these features are:

\begin{itemize}

  \item   It should yield a convergent  path integral 

  \item   The resulting wavefunction should predict  classical
  spacetime in the late universe, i.e, oscillating behaviour when the $\Psi $ is well approximated by the semiclassical approximation.

   \item  The resulting wavefunction should obey the diffeo\-mor\-phism con\-straints, in par\-ticular the Wheeler-DeWitt equation.

\end{itemize}

It is well known that any integration contour over real metrics would yield a wavefunction that does not satisfy any of these basic requirements. On the other hand, an integration contour over complex metrics can, if wisely chosen, lead to a wavefunction that satisfies those requirements. 

In the simplicial framework, complex metrics arise from complex-valued
squared edge lengths, $(\ref{simpmet})$. The boundary squared edge
length, $S$, has to be real and positive for obvious physical reasons. But the interior squared edge length, $\xi $, can be allowed to take complex values. Before we attempt to choose the correct integration contour it is essential we study the analytical and asymptotic properties of the action as a function of the complex variable $\xi $.

\subsection{Analytic Study of the Action}

By allowing $\xi$ to be complex-valued we have to contend with a
multivalued   Regge action, $(\ref{actionI} )$. The deficit angle terms
have an infinite number of branches, and we must deal with this
multivaluedness, if we are to study the behaviour of steepest descents
contours,(along which the imaginary part of the action must remain
constant). Furthermore, in order to obtain a continuous and
single-valued action, a careful analysis of all the action's 
branching points is necessary. As expected these branching points occur
at the values of $\xi$ for which the simplices become degenerate, i.e,
where their volumes vanish. The  analytic study of the action
  should then enable us 
to  obtain integration contours that avoid such branching points, and
along which the action is continuous and single-valued, leading to a
well defined wavefunction of the Universe.

The action $(\ref{actionI} )$ is trivially analytic with respect to
the variables $\phi _{i},\;\phi _{b}$ and $S$. But its dependence on
the complex-valued $\xi $ is much more complicated. So we shall investigate the analytic properties of $I$ as if it was a function of $\xi $ only, $I=I[\xi ]$, the other variables acting as parameters.

The function $I[\xi ]$ has singularities at $\xi=0$ and $\xi=1/3$, and
square root branch points at $\xi
=1/4,\:1/3$ and $3/8$. These branch points correspond respectively to the vanishing of the volume of the internal $2-$simplices,  $3-$simplices and $4-$simplices. Using 
$$\arccos{z}=-i\log (z+\sqrt{z^{2}-1})$$

we see that $\xi =1/3$ is also a logarithmic branch point, near which
the action behaves like :

\be
I\:\sim i2\sqrt{3}N_{3}\biggl(\frac{S}{H^{2}}\biggr)\log{(3\xi -1)}
\ee

The multivaluedness of $I[\xi ]$ associated with these branch points
forces us to implement branch cuts in order to obtain a continuous
function. In general, for terms of the type $\sqrt{z-z_{0}}$ we consider a branch cut $(-\infty ,z_{0}]$. So the branch cuts associated with the terms $\sqrt{\xi -3/8}$, $\sqrt{\xi -1/4}$ and $\sqrt{\xi -1/3}$, altogether lead to a branch cut  $(-\infty ,3/8]$. On the other hand, terms of the type $\arccos (z)$ have branch points at $z=-1,+1,\infty $, and usually the associated branch cuts are chosen as $(-\infty ,-1]\cup [1,+\infty )$. These terms are also infinitely multivalued.

The corresponding cuts for the term $\arccos{\frac{2\xi-1}{6\xi -2}}$
are $(\frac{1}{3},\frac{3}{8}]\cup \-[\frac{1}{4},\frac{1}{3})$. On
the other hand, associated with the term $\arccos{\frac{1}{2\sqrt{6\xi -2}}}$ we have one cut  $(\frac{1}{3},\frac{3}{8}]$ associated with $\arccos{u(z)}$, and another $(-\infty, \frac{1}{3}]$ associated with $u(z)=\sqrt{6\xi -2}$.

So when we consider all these branch cuts simultaneously, we see that one way to obtain a continuous action $I$ as a function of $\xi $, is to consider a total branch cut $(-\infty ,\frac{3}{8}]$. Note that this also takes care of the singularity at $\xi =0$

Although the action  then becomes a continuous function of $\xi $ in
the complex plane with a cut $(-\infty ,\frac{3}{8}]$, it is still
infinitely multivalued. As usual in similar cases in order to remove
this multivaluedness we redefine the domain where the action is
defined, from the complex plane to the Riemann surface associated with
$I$. The infinite multivaluedness of the action reflects itself in $I$
having an infinite number of branches with different values. The Riemann surface is composed of an infinite number of identical sheets, $\bkC -(-\infty ,\frac{3}{8}]$, one sheet for each branch of $I$.  

We define the first sheet ${\bkC} _{1}$ of $I[\xi ]$ as the sheet
where the terms in $\arccos (z)$ assume their principal  values. So
the action in the first sheet will be formally equal to the original
expression $(\ref{actionI} )$, with positive signs taken for the
square root factors. Note that with the first sheet defined in this way, for real $\xi >3/8$ the volumes and deficit angles are all real, leading to a real Euclidean  action for $\xi \in [\frac{3}{8},+\infty )$ on the first sheet. 
On the other hand, when $\xi $ is real and less than $1/4$ in the first sheet , the volumes become pure imaginary and the Euclidean action becomes pure imaginary. For all other points of this first sheet the action is fully complex.

Since by $(\ref{simpmet} )$ we see that the simplicial metric in each
$4-$simplex is real $iff$ $\xi $ is real, then the simplicial
geometries built out of these $4-$simplices will be real when  $\xi $
is real. Furthermore the corresponding eigenvalues of $g_{ij}$ are
$\lambda =\{4(\xi -3/8),1/2,1/2,1/2\}$, \cite{birm}. So we see that for real $\xi >3/8$ we have real Euclidean signature geometries, with real Euclidean action, and for real $\xi <1/4$, we have real Lorentzian signature geometries with pure imaginary Euclidean action.

Up to now we have been concerned with what happens in the first sheet only. When we continue the action in $\xi $ around one or more branch points, we will leave this first sheet and emerge in some other sheet of the Riemann surface. Only a few of these other sheets are relevant to us.

When the action is continued in $\xi $ once around all finite branch points ($\xi = 1/4,1/3,3/8$), we reach what shall be called the second sheet . It is easy to conclude that the action in this second sheet is just the negative of the action in the first sheet.

\be
I^{I}[\xi ,S,\phi _{i},\phi _{b}]=-I^{II}[\xi ,S,\phi _{i},\phi _{b}] 
\ee

Once in the second sheet, if we encircle the branch points in such a way that we cross the branch cut, $(-\infty ,\frac{3}{8}]$, between $1/4$ and $1/3$, we arrive at what we shall call the third sheet. By doing this the terms $\pi -2\arccos{\frac{1}{2\sqrt{6\xi -2}}}$ and $\arccos{\frac{2\xi -1}{6\xi -2}}$, both flip signs and so the action in this third sheet is 

\begin{eqnarray*}
I^{III}[\xi ,S,\phi _{i},\phi _{b}]&=&-\frac{S}{H^{2}} \biggl\{ \biggl(N_{3}\sqrt{3}\biggr)\biggl[\pi -2\arccos {\frac{1}{2\sqrt{6\xi -2}}}\biggr] \\
                                   &-&N_{1}\sqrt{\xi -1/4}\biggl[2\pi +\frac{6N_{3}}{N_{1}}\arccos{\frac{2\xi -1}{6\xi -2}}\biggr]  \\
                                   &-& \biggl(\frac{N_{3}}{120\sqrt{2}}\biggr)\frac{\sqrt{\xi -3/8}}{\xi}(\phi _{i}l-\phi _{b}l)^{2}\biggr\} +\frac{S^{2}}{H^{2}}\biggl\{\biggl(\frac{N_{3}}{4\sqrt{2}}\biggr)\sqrt{\xi -3/8}  \\  
                                   &+& \mbox{} \frac{N_{3}}{240\sqrt{2}}\biggl(\frac{m^{2}l^{2}}{H^{2}}\biggr)\sqrt{\xi -3/8}(\phi _{i}^{2}l^{2}+4\phi _{b}^{2}l^{2})\biggr\}
\end{eqnarray*}

If instead of crossing the cut between $1/4$ and $1/3$, we transverse
it between $1/3$ and $3/8$ then we will end up in a different
sheet. However, the asymptotic behaviour of the action is similar to
that on the third sheet, so we will not go into the details.

\subsection{Asymptotic Behaviour of the Action}

In any investigation of the correct contour of integration it is
essential to know how the action behaves asymptotically with respect
to the variable $\xi , \: (\xi \rightarrow \infty )$, because only then will we be able to evaluate the contribution coming from these regions. 

In the first sheet when $ \xi \rightarrow \infty $ the action behaves like

\be
I^{I}[\xi ,S,\phi _{i},\phi _{b}]\sim \frac {\frac{N_{3}}{4\sqrt{2}}+A(\phi _{i},\phi _{b})}{H^{2}} S(S-S_{crit})\sqrt{\xi } 
\ee
where 

\be 
A(\phi _{i},\phi _{b})=\frac{1}{240\sqrt{2}}\frac{m^{2}l^{2}}{H^{2}}N_{3}(\phi _{i}^{2}l^{2}+4\phi _{b}^{2}l^{2}) 
\ee
and

\be
S_{crit}=\frac{N_{1}[2\pi -\frac{6N_{3}}{N_{1}}\arccos {(1/3)}]}{\frac{N_{3}}{4\sqrt{2}}+ A(\phi _{i},\phi _{b})}  
\ee

The asymptotic behaviour of $I^{I}$ for large $\xi $ depends on whether or not $S$ is larger than the critical value $S_{crit}$. This value will also be crucial for the classical solutions to be obtained in the next section.

First note that there are two important special cases:

\begin{itemize}

\item  The pure gravity model, $m=0, \phi =0$, in which we are reduced to the kind of models considered in \cite{birm}.

\item  The massless scalar field model, $m=0$.

\end{itemize}

In both cases $ A(\phi _{i},\phi _{b})=0$, and the asymptotic behaviour becomes independent of the matter field sector. In both cases as  $ \xi \rightarrow \infty $ we have:

\be 
I^{I \:m=0}[\xi ,S,\phi _{i},\phi _{b}]\sim \frac{N_{3}}{4\sqrt{2}}\frac{1}{H^{2}}S(S-S_{crit}^{m=0})\sqrt{\xi }  
\ee
where

\be
S_{crit}^{m=0}=\frac{4\sqrt{2}N_{1}}{N_{3}}\biggl[2\pi -\frac{6N_{3}}{N_{1}}\arccos {(1/3)}\biggr]  
\ee

Note that once $A(\phi _{i},\phi _{b})\geq 0$, then $ S_{crit}^{m=0}\geq S_{crit}$.

For the second sheet obviously the asymptotic behaviour of the action
is just the negative of that in the first sheet. For the third sheet
the situation is slightly different because only some terms in the
action change sign and we have that as $ \xi \rightarrow \infty $:

\be
I^{III}[\xi ,S,\phi _{i},\phi _{b}]\sim \frac {\frac{N_{3}}{4\sqrt{2}}+A(\phi _{i},\phi _{b})}{H^{2}} S(S+S^{III}_{crit})\sqrt{\xi } 
\ee
where 

\be
S^{III}_{crit}=\frac{N_{1}[2\pi +\frac{6N_{3}}{N_{1}}\arccos {(1/3)}
]}{\frac{N_{3}}{4\sqrt{2}}+ A(\phi _{i},\phi _{b})}  
\ee

As before we consider the two sub-cases where $ A(\phi _{i},\phi
_{b})=0$

\be 
I^{III \:m=0}[\xi ,S,\phi _{i},\phi _{b}]\sim
\frac{N_{3}}{4\sqrt{2}}\frac{1}{H^{2}}S(S+S_{crit}^{III  m=0})\sqrt{\xi }  
\ee
and

\be
S_{crit}^{III m=0}=\frac{4\sqrt{2}N_{1}}{N_{3}}\biggl[2\pi +\frac{6N_{3}}{N_{1}}\arccos {(1/3)}\biggr]  
\ee

\sect{Classical Solutions}

The classical simplicial geometries are the extrema of the Regge action we have obtained above. In our minisuperspace model there are two degrees of freedom $\xi ,\phi _{i}$. So the Regge equations of motion will be:

\be
\frac{\partial I}{\partial \xi}=0
\label{classic1}
\ee
and

\be
\frac{\partial I}{\partial \phi _{i}}=0
\label{cla2}
\ee

They are to be solved for the values of $\xi ,\phi _{i}$, subject to
the fixed boundary data $S, \phi _{b}$. The classical solutions will
thus be of the form $\overline{\xi}(S, \phi _{b})$, and
$\overline{\phi} _{i}(S,\phi _{b})$. The solution $\overline{\xi}(S,
\phi _{b})$ completely determines the simplicial geometry. For the
general model equations$(\ref{classic1} )$, $(\ref{cla2} )$, take the form

\be
\label{cleqS}
S=\frac{  \frac{a_{1}}{2}\frac{1}{\sqrt{\xi -1/4}}[2\pi -a_{2}\arccos{\frac{2\xi -1}{6\xi -2}}]+\frac{N_{3}}{240\sqrt{2}}\frac{\xi -3/4}{\xi ^{2}\sqrt{\xi -3/8}}(\phi _{i}-\phi _{b})^{2}l^{2}}{ \frac{a_{3}}{2\sqrt{\xi -3/8}}+\frac{1}{480\sqrt{2}}(\frac{m^{2}l^{2}}{H^{2}})\frac{N_{3}}{\sqrt{\xi -3/8}}({\phi }_{i}^{2}l^{2}+4{\phi} _{b}^{2}l^{2})}  
\ee
and

\be
\label{cleqphi}
\phi _{i}=\frac{\phi _{b}}{1+\frac{1}{2}\frac{m^{2}l^{2}}{H^{2}}\xi S}  
\ee

Note that we will be expressing  the values of the field $\phi $ and its
mass $m$, in Planck units ($l^{-1}$). Furthermore, for simplicity of
notation we have introduced  $a_{1}=N_{1}$, $a_{2}=6N_{3}/N_{1}$ and $a_{3}=N_{3}/(4\sqrt{2})$.

Note that these expressions were obtained using the expression for the action on the first sheet. Of course, since on the second sheet the action is just the negative of this,  the equations of motion are the same. And obviously every classical solution   $\overline{\xi}_{I}(S, \phi _{b})$ located on the first sheet will have a counterpart $\overline{\xi}_{II}$ of the same numerical  value, but located on the second sheet, and so with an action of opposite sign, $I[\overline{\xi}_{I}(S, \phi _{b})]=-I[\overline{\xi}_{II}(S, \phi _{b})]$. So the classical solutions occur in pairs.

For physical reasons the squared boundary edge length, $S$, has to be
real and  positive, and  we will also assume that $\phi _{b}$ is real.  

Before we go on to study the general model it is useful to investigate
the two special cases mentioned in section $\bf{3.6}$, i.e., pure gravity, and massless scalar field.

\subsection{Pure Gravity}

In this case our model is reduced to a family of models already
considered by \cite{birm}, where we have only one metric degree of freedom $\xi $ and its associated equation of motion is 

\be
S=\frac{a_{1}}{a_{3}}\sqrt{\frac{\xi -3/8}{\xi -1/4}}\biggl[2\pi -a_{2}\arccos{\frac{2\xi -1}{6\xi -2}}\biggr]
\ee

The condition of real positive $S$ means that physically acceptable
classical solutions occur only for real $\xi >3/8$ and real $\xi
<1/4$. There are two different kinds of classical solutions according
to whether $S_{crit}^{m=0}$ is negative or positive. 

  For positive $S_{crit}^{m=0}$ the general form of the solutions is
  similar to that of the ${\cal S}^{2}\times {\cal S}^{1}$ model in Figure $1$.There are two different regimes: For every $S$ between $0$ and $S_{crit}$, the classical solutions ${\overline{\xi }}_{I}(S)={\overline{\xi }}_{II}(S)$ are real and belong to $[3/8,+\infty )$. So they correspond to Euclidean signature simplicial geometries and their Euclidean actions will be real (but symmetric). For every $S>S_{crit}$ the classical solutions ${\overline{\xi }}_{I}(S)={\overline{\xi }}_{II}(S)$ are real and belong to $(-\infty , 1/4]$. So they correspond to Lorentzian signature simplicial geometries  and their Euclidean actions are pure imaginary

\be
\label{Loraction}
I[\overline{\xi }_{I}(S)]=-I[\overline{\xi }_{II}(S)]=i\tilde{I}[\overline{\xi }_{I}(S)] \ee
where $\tilde{I}[\overline{\xi }_{I}(S)] $ is real.

 When $S_{crit}$ is negative  there is no Euclidean regime. All classical solutions are Lorentzian geometries. For each $S>0$, there is a pair of solutions ${\overline{\xi }}_{I}(S)={\overline{\xi }}_{II}(S)\in (-\infty ,1/4]$ on the first and second sheets. Their corresponding actions are pure imaginary, as in $(\ref{Loraction}  )$.

\subsection{Massless Scalar Field}

In this case, as in the general model there are two degrees of freedom  
$\xi $ and $\phi _{i}$. The corresponding equations of motion are

\be
S=\frac{a_{1}}{a_{3}}\sqrt{\frac{\xi -3/8}{\xi -1/4}}\biggl[2\pi -a_{2}\arccos{\frac{2\xi -1}{6\xi -2}}\biggr]+\frac{N_{3}}{120\sqrt{2}a_{3}}\frac{1}{\xi}\biggl(1-\frac{3}{4\xi}\biggr)(\phi _{i}-\phi _{b})^{2}  
\ee
and

\be
\phi _{i}=\phi _{b}  
\ee

If we substitute the second equation into the first we see that it
reduces to the equation for $\xi $ of the previous case. So the
classical solutions for $\xi $ in the massless case coincide with the
pure gravity case, $\overline{\xi }(S,\overline{\phi }_{i}=\phi
_{b})=\overline{\xi }(S)$. So the presence of the massless scalar field
does not change the simplicial geometries that are classical
solutions. Furthermore, the actions associated with these classical
solutions are independent of the scalar field and coincide with the
actions associated with the previous model.

\be
I[{\overline{\xi }},{\overline{\phi}}_{i}=\phi _{b}]=I[{\overline{\xi }}]
\ee

\subsection{Massive Scalar Field}

Given the classical equations of motion $(\ref{cleqS})$ and
$(\ref{cleqphi} )$, introducing the second equation into the first
yields a cubic equation on $S$ for each value of $\xi$, given fixed
$\phi _{b}$. That equation is 

\be
\label{cleq}
A_{3}(\xi)S^{3}+A_{2}(\xi)S^{2}+A_{1}(\xi)S+A_{0}(\xi)=0
\ee
where

\be
A_{3}(\xi)=\frac{a_{3}}{2}\biggl[K^{2}+\frac{2}{15}K^{3}\phi_{b}^{2}\biggr]\xi^{2} 
\ee

\begin{eqnarray*}
\label{cleqC}
A_{2}(\xi)&=& a_{3}\biggl[K+\frac{2}{15}K^{2}\phi_{b}^{2}\biggr]\xi
-\frac{a_{1}}{2}K^{2}\xi^{2}\sqrt{\frac{\xi-3/8}{\xi-1/4}}\biggl[2\pi-a_{2}\arccos{\frac{2\xi
-1}{6\xi -2}}\biggr] \\
          &-&\mbox{}  \frac{a_{3}}{60}K^{2}\phi_{b}^{2}(\xi-3/4) 
\end{eqnarray*}

\be
A_{1}(\xi)=\frac{a_{3}}{2}\biggl(1+\frac{K\phi_{b}^{2}}{6}\biggr)-a_{1}K\xi\sqrt{\frac{\xi-3/8}{\xi-1/4}}\biggl[2\pi-a_{2}\arccos{\frac{2\xi
-1}{6\xi -2}}\biggr]
\ee

\be
A_{0}=-\frac{a_{1}}{2}\sqrt{\frac{\xi-3/8}{\xi-1/4}}[2\pi-a_{2}\arccos{\frac{2\xi -1}{6\xi -2}}]
\ee
where $K=(1/2)(ml/H)^{2}$.

The more complicated dependence on $\xi$ leads us to solve the
classical equation in order to find $S$ for each value of $\xi$. 
By inverting the resulting solutions we  obtain
$\overline{\xi }(S,\phi _{b})$, and consequently $\overline{\phi
}_{i}(S,\phi _{b})$, via $( \ref{cleqphi} )$. This being a cubic
equation we expect 3 solutions for each value of $\xi$. However,
because of obvious physical constraints we will accept only solutions
that are real and positive.

 Of course as in the pure gravity model, classical solutions occur in pairs, one on the first sheet and the other on the second sheet. They have the same numerical value but yield actions of opposite sign.

Instead of presenting $\overline{\xi}=\overline{\xi }(S,\phi _{b})$ 
as a $3-$D plot it is more informative to consider the graphs of
$\overline{\xi}=\overline{\xi }(S)$ for several values of $\phi
_{b}$. Figure $2$ shows one such plot for a cone over the $\alpha
_{4}$ triangulation of the $3-$sphere, in the presence of a scalar field of mass
$m=1(l^{-1})$, and a boundary value $\phi _{b}=1 (l^{-1})$.


 Note that  near $\xi =1/4$ the solution is similar to a solution in the pure gravity model. This is to be expected because around $\xi =1/4$ the curvature (Ricci scalar) term dominates all others, including the matter terms in the action.  So for large values of $S$, that is, for $\xi \rightarrow \frac{1}{4}^{-}$, the classical solutions in the massive scalar field model will be well approximated by their analogues in the pure gravity model.

Away from the singularity the situation is different. In the Euclidean
regime, $\xi>3/8$, there is always only one real positive solution of
$(\ref{cleq})$ for each value of $\xi$. Furthermore,  as $\xi \rightarrow +\infty$,  $S$ converges to a critical value dependent on the
value of the mass $m$ and boundary value of the scalar field $\phi
_{b}$:

\be
S_{crit}^{m}=\frac{a_{1}}{a_{3}}\biggl[2\pi
-a_{2}\arccos
{(1/3)}\biggr]\biggl(\frac{1}{1+\frac{2}{15}K\phi_{b}^{2}}\biggr)=S_{crit}^{m=0}\times\biggl(\frac{1}{1+\frac{2}{15}K\phi_{b}^{2}}\biggr)
\ee

This way the effective critical value decreases as the values of $m$
or $\phi_{b}$ increase. 

In the Lorentzian regime, $\xi<1/4$, the presence of matter radically
changes the classical solutions. When $\xi$ is sufficiently negative, i.e.,
$\xi <\xi_{0}$,
there are $3$ real positive solutions of $(\ref{cleq})$ for each value of
$\xi$. One of these branches is almost constant and  rapidly
converges to   $S_{crit}$ as $\xi$ decreases. Another  branch also
starts at $\xi _{0}$, but then  converges to
zero as $\xi \rightarrow -\infty$. Finally, the third branch,
which also 
converges to zero as $\xi \rightarrow -\infty$, but  continues all
the way to $\xi=1/4$, where $S\rightarrow +\infty$. It is this
last branch of classical solutions that is relevant to us, because
it is the only one that continues beyond $S_{crit}$ all the way to
$S\rightarrow +\infty$.

 So for $S\in [0,S_{crit})$, we will have:

\begin{itemize}

\item  Two pairs of real Lorentzian signature solutions $\overline{\xi
}_{I}^{L1}(S,\phi _{b})=\overline{\xi }_{II}^{L1}(S,\phi _{b}) \in
(-\infty ,1/4]$, and $\overline{\xi }_{I}^{L2}(S,\phi _{b})=\overline{\xi }_{II}^{L2}(S,\phi _{b}) \in (-\infty ,1/4]$ with pure imaginary Euclidean actions.

\item One pair of real Euclidean signature solutions $\overline{\xi
}_{I}^{E}(S,\phi _{b})=\overline{\xi }_{II}^{E}(S,\phi _{b}) \in
[3/8,+\infty )$,  with real Euclidean action.

\end{itemize}

 For $S >S_{crit}$, we will have:

\begin{itemize}

\item  Only one pair of real solutions $\overline{\xi }_{I}(S,\phi _{b})=\overline{\xi }_{II}(S,\phi _{b}) \in (-\infty ,1/4]$ that correspond to Lorentzian signature simplicial metrics, and whose Euclidean actions, though symmetric, are both pure imaginary. 

\end{itemize}

$$ I[\overline{\xi }_{I}(S,\phi _{b})]=-I[\overline{\xi }_{II}(S,\phi _{b})]=i\tilde{I}[\overline{\xi }_{I}(S,\phi _{b})]$$

If we increase the value of $m$ or $\phi _{b}$, the value of $S_{crit}$
decreases to zero and eventually the branch associated with the Euclidean
regime vanishes. The same happens to the two extra Lorentzian
branches that are contained in  $[0,S_{crit})$, and we are left with a situation similar to the one
we found in the pure gravity model when $S_{crit}^{m=0}$ becomes $<0$. That
is, there remains only one Lorentzian branch for the entire range
of $S$, from $S=0$ to $S\rightarrow +\infty $. For other topologies
like ${\cal T}^{3}$ there is no Euclidean regime whatever the value
of the field, because $S_{crit}<S_{crit}^{m=0}<0$. See Figure $3$.


\sect{Semiclassical Approximation}

One of the main requirements on any model is that it yields a wavefunction that in the late universe predicts a classical (Lorentzian) spacetime, like the one we experience. Now, a wavefunction of the universe will predict a classical spacetime where it is well approximated by the semiclassical approximation associated with Lorentzian classical solutions. We shall see what are the conditions that lead to such a situation in the next section. For now we will assume that this is the case and concentrate on the semiclassical approximation.

The semiclassical wavefunction for our model will be obtained from the full wavefunction

\be
\Psi (S,\phi _{b})=\frac{S}{2\pi iH^{2}}\int _{C}d\xi d\phi _{i}e^{-I(\xi,S,\phi _{i},\phi_{b})}
\ee

by assuming that the integral is dominated by the contributions of the
stationary points of the Regge action, i.e., the classical solutions computed above.

We shall consider that the first integration is over the complex
valued $\xi$ and then over the real valued $\phi _{b}$.
We then  assume that for each pair of boundary data $(S,\phi _{b})$,
the integral
 over $\xi$ is  dominated by the
contributions coming from   the classical
solutions $\{\overline{\xi}_{k}(S,\phi _{b})\}$. If these  are
real Lorentzian
 solutions  with purely imaginary actions
$I_{k}=i\tilde{I}[\overline{\xi}_{k}(S,\phi _{b});\phi
_{i}]=i\tilde{I}_{k}(S,\phi _{b},\phi _{i})$, then the semiclassical
approximation for the integral over $\xi $ is

\be
\int _{C}d\xi e^{-I(\xi,S,\phi _{i},\phi_{b})}\sim 
\sum_{k}\sqrt{\frac{S^{2}}{2\pi H^{4}\mid
\tilde{I}^{''}[\overline{\xi}_{k}(S,\phi _{b}),\phi _{i}]\mid }}e^{-i[
\tilde{I}(\overline{\xi}_{k}(S,\phi _{b}),\phi _{i})  +\mu _{k}\frac{\pi}{4}]}
\ee
where $'$ means derivative with respect to $\xi $
and $\mu _{k}=sgn ( \tilde{I}^{''})$.

The remaining integrals over $\phi _{i}$ are now  Fourier-type
integrals, and can be evaluated by the stationary phase method
by the contribution coming from their stationary points which are
precisely the classical solutions $\overline{\phi}_{i}^{k}(S,\phi _{b})$

\begin{eqnarray*}
\Psi _{SC}(S,\phi _{b})&\sim &\int d\phi _{i}\sum_{k}\sqrt{\frac{S^{2}}{2\pi H^{4}\mid
\tilde{I}^{''}[\overline{\xi}_{k}(S,\phi _{b}),\phi _{i}]\mid }}e^{-i[
\tilde{I}(\overline{\xi}_{k}(S,\phi _{b}),\phi _{i})  +\mu _{k}\frac{\pi}{4}]} \\
                       &\sim & \mbox{} \sum_{k}\sqrt{\frac{S^{2}}{2\pi
H^{4}\mid \tilde{I}_{k}^{''}(S,\phi _{b})\mid }}e^{-i[
\tilde{I}_{k}(S,\phi _{b})  +\mu _{k}\frac{\pi}{4}]}
\label{prefac}
\end{eqnarray*}
where $I_{k}(S,\phi _{b})=I[\overline{\xi}_{k}(S,\phi _{b}),\overline{\phi}_{i}^{k}(S,\phi _{b})]$.

When the dominant extrema are real Euclidean solutions
$\{\overline{\xi}_{k}(S,\phi _{b})\}$,  with real Euclidean
 actions $I_{k}(S,\phi _{b})$  then the semiclassical
evaluation of the integral over $\xi $ leads to Laplace-type integrals
over $\phi _{i}$ which once more are dominated by the contributions
coming from the stationary points of $I[\overline{\xi}_{k}(S,\phi
_{b}),\phi _{i}]$,
which are precisely the classical solutions
$\overline{\phi}_{i}^{k}(S,\phi _{b})$.
So the semiclassical wavefunction will then be

\be
\Psi _{SC}(S,\phi _{b})\sim \sum_{k}\sqrt{\frac{S^{2}}{2\pi
H^{4}\mid I_{k}^{''}(S,\phi _{b})\mid }}e^{-
I_{k}(S,\phi _{b})}
\ee

We shall be mainly interested in the behaviour of the wavefunction of
the universe for
 large $S$, (relative to  Planck`s length, $l$) because that
corresponds to the late universe we experience.

The semiclassical approximations associated with pure gravity models
were calculated in \cite{h2}  and \cite{birm} . In these
 models when $S$ is large, the only classical solutions are a pair of
real Lorentzian geometries,  
${\overline{\xi }}_{I}(S)={\overline{\xi }}_{II}(S)\in (-\infty
,1/4]$, with purely imaginary Euclidean actions,
 and so the semiclassical approximation associated with them
yields an oscillating wavefunction for $S>S_{crit}$, as desired.


For the massless scalar field model the classical solutions have
the same  simplicial geometry, as  in pure gravity case, since
 ${\overline{\phi _{i} }}=\phi _{b}$, implies that
 ${\overline{\xi }}_{k}(S,\phi_{b})={\overline{\xi }}_{k}(S)$,
 where $k=I,II$. So the action
associated with the classical solutions is independent of $\phi
_{i}$,  and the influence of the scalar field is visible only through 
the pre factor in $(\ref{prefac})$.


In the general model, with a massive scalar field, we have seen that in our
range of interest, that is, large $S$, there is also a pair of real
Lorentzian classical solutions, for each value of $S$ (and $ \phi
_{b}$).
 The
semiclassical wavefunction associated with both solutions has an
oscillating behaviour as required in order to predict classical
spacetime for the late universe. In Figures $4$,$5$, and $6$ we
show the numerically calculated 
 semiclassical wavefunctions associated with simplicial geometries 
that are cones over triangulations of some of the most
relevant  spatial $3-D$ universes, namely simple triangulations of
the $3-$sphere $S^{3}$, of  $S^{2}\times S^{1}$, and $T^{3}$,
 in the presence of massive scalar fields. The
results clearly indicate an oscillatory behaviour for large $S$,
signalling  classical spacetime, in all cases. In the case of $T^{3}$,
 $S_{crit}<0$, which means that the semiclassical wavefunction is
valid for
all values of $S$, and not just large ones. Similar calculations
for other triangulations of the spatial $3-D$ universe yield similar 
wavefunctions, confirming the generality of the result.


\sect{Steepest Descents Contours}

We will now investigate under what conditions the semiclassical approximation 
is a good approximation of the full wavefunction. Mathematically this happens when the contour of integration $C$ can be distorted such that it passes as a steepest descent contour $C_{SD}$ through the classical solutions (on which we wish to base our semiclassical approximation) so that the integral is dominated by the contribution from the neighbourhoods of those solutions.

In order to guarantee that this is indeed the case, a detailed knowledge of the entire contour of integration is not necessary. It is sufficient to have  such  a knowledge near  the extrema that supposedly dominate the integral, and prove that the integrand is sharply peaked around the classical solutions and that there are no other ``critical points'' yielding significant contributions to the integral. 

In our model there are two integration variables $\xi $ and $\phi _{i}$ and the full wavefunction of the universe  is given by $(\ref{ppsi} )$. We work under the assumption that $\phi _{i}$ is to be integrated over real values, and $\xi $ over the complex Riemann surface of the Euclidean action $I$. In order to justify the validity of the  semiclassical approximation for such wavefunction we need to prove that the integral over $\xi $, can be calculated as a steepest descent $(SD)$ integral for all the relevant values of $\phi _{i}$. 

Assuming that this is true, we can replace $\int _{C_{SD}}d\xi e^{-I}$ by its   semiclassical approximation based on the relevant classical extrema. This will give rise to Laplace type integrals in $\phi _{i}$, when the extrema have real Euclidean action, and to Fourier type integrals in $\phi _{i}$, when the extrema have pure imaginary Euclidean action. These integrals can  then be shown to be dominated by the stationary points of the integrand which coincide with the classical solutions $\overline{\phi}_{i}$, where 

$$ \frac{\partial I[\overline{\xi},\phi _{i}]}{\partial \phi _{i}}\mid _{\phi _{i}=\overline{\phi}_{i}}=0 $$   

This justifies the validity of the semiclassical approximation to the full wavefunction.

In general, a SD contour associated with an extremum ends up either at
$\infty $ or at a singular point of the integrand, or at another
extremum with the same value of $Im(I)$. We have seen that in all
three models considered, i.e., pure gravity, massless scalar field,
and massive scalar field, when $S$ is big enough the only classical
solutions are a pair or real Lorentzian solutions $(\overline{\xi
}_{I}(S,\phi _{b}),\overline{\phi }_{i}(S,\phi _{b}))$ and
$(\overline{\xi }_{II}(S,\phi _{b}),\overline{\phi }_{i}(S,\phi
_{b}))$, where  $\overline{\xi }_{I}=\overline{\xi }_{II}<1/4$. They
are located on the first and second sheets respectively, and so have
pure imaginary actions of opposite sign. Given that their actions are
different valued no single SD path can go directly from one to the other
extremum. On the other hand given that

$$ I[\overline{\xi},\phi _{b}]=[I[\overline{\xi}^{*},\phi _{b}]]^{*} $$

and $$ I[\overline{\xi}_{I},\phi _{b}]=- I[\overline{\xi}_{II},\phi _{b}]$$
where $*$ denotes complex conjugation, we see that the SD path that passes through  $\overline{\xi }_{II}$ will be
the complex conjugate of the SD path that passes through $\overline{\xi }_{I}$.
So the total SD contour will always be composed of two complex
conjugate sections, each passing through one extremum, and this
together with the real analyticity of the action guarantees that the
resulting wavefunction is real.

\subsection{Pure Gravity Model}

In the pure gravity model the only singularity is located at $\xi
=1/3$, where $Im(I)$ diverges and so no SD contour that passes through
a classical solution can end up at such a singularity. So in this model the SD paths passing through both classical Lorentzian extrema are condemned to end up at infinity. So we only have to worry about convergence of the integral when $\xi \rightarrow \infty $.

The appropriate SD contour for $S>S_{crit}$

\be
C_{SD}(S)=\biggl\{\xi \in R : Im[I(S,\xi )]=\tilde{I}[\overline{\xi}(S)]\biggr\}
\ee
(where $R$ is the Riemann sheet of the action), can be shown to be
 made up of two sections, one passing through the Lorentzian
 extremum in the first sheet and the other its complex conjugate, passing  through the extremum 
located in the second sheet.

For the SD path associated with $\overline{\xi }_{I}$ starting off in the first sheet  the SD contour on the upper half of this sheet is given asymptotically by

\be
\frac{a_{3}}{H^{2}}S(S-S_{crit})Im(\sqrt{\xi})=\tilde{I}[\overline{\xi}_{I}] 
\ee
and we can guarantee the convergence of the integral along this part of the SD contour because the real part of the action is asymptotically

\be
Re [I(\xi ,S)]\sim \frac{a_{3}}{H^{2}}S(S-S_{crit}^{m=0})\sqrt{\mid \xi \mid}
\ee
The SD contour then passes through $\overline{\xi }_{I}\in
(-\infty ,1/4]$ crossing into the second sheet of the
action. Proceeding along the SD contour we will then transverse the
branch cut $(-\infty ,3/8]$, again. When $S$
is sufficiently large,  the crossing happens between $1/4$ and $1/3$,
and we emerge in the third sheet. If $S$ is smaller, the crossing happens
between $1/3$ and $3/8$, and we emerge in a different sheet, but the
asymptotic behaviour of the action is similar and so in terms of the
convergence of the integral, the two situations are equivalent. In the
 third sheet the SD contour then proceeds to infinity in the first
 quadrant along the curve that is asymptotically defined by

\be
\frac{a_{3}}{H^{2}}S(S+S^{III m=0}_{crit})Im(\sqrt{\xi})=\tilde{I}[\overline{\xi}_{I}] 
\ee

Once again the convergence is guaranteed by the asymptotic behaviour of the action

\be
Re [I(\xi ,S)]\sim \frac{a_{3}}{H^{2}}S(S+S_{crit}^{III m=0})\sqrt{\mid \xi \mid}
\ee

Figure $7$ shows a numerical computation of such an SD contour
for the case of ${\cal M}^{3}=T^{3}$.


The asymptotic behaviours of the two sections are such that  
they meet  at infinity  on the first and third
sheets, thus obtaining a closed SD contour defining the wavefunction
of the universe.

The SD integral over $\xi $ is thus  well approximated by the
semiclassical approximation associated with these two Lorentzian
extrema and so it yields a wavefunction that in the late universe
(large $S$) predicts classical Lorentzian spacetime that is a 
solution of the Einstein GR equations, as desired.

\subsection{Massless Scalar Field model}

In the massless scalar field model, the situation is somewhat different. There is a second integration on the variable $\phi _{i}$. So if in the previous case we had a family of SD contours labelled by the value of $S$ we now have a family of SD contours labelled by two variables, namely $S$ and $\phi _{i}-\phi _{b}$.

\be
C_{SD}(S,\phi _{i}-\phi _{b})=\biggl\{\xi \in R : Im[I(S,\xi ,\phi
_{i}-\phi _{b})]=\tilde{I}[\overline{\xi}(S,\phi
_{b}),\overline{\phi}_{i}=\phi _{b})]\biggr\}
\ee
where $R$ is the Riemann sheet of the action.

We must be able to obtain the SD contour in $\xi $ for each value of
$S$ and $\phi _{i}-\phi _{b}$, and prove that it verifies the
conditions mentioned above for all relevant values of  $\phi _{i}$ and
$S$. The numerical computation of these steepest descent contours, for
values $S>S_{crit}^{III m=0}$, 
yields two different kinds of contours. For large enough values of $S$,
the SD contours for each value of $\phi _{i}-\phi _{b}$, are similar
to the SD contours we have encountered in the pure gravity model. See Figure $8$.


The fact that the asymptotic behaviour of the action is the same in
both the pure gravity model and the massless scalar field model
guarantees the convergence of the SD contour.

However, when the value of $S$ is smaller the influence of the
 singularity at $\xi =0$ dominates and we obtain a different kind of
 SD contour. In Figure $9$ we show the SD contour for a cone over the
 $3-$sphere triangulation $\alpha _{4}$, passing through the classical
 Lorentzian solution located in the first sheet,
 $\overline{\xi }_{I}$, when $S=50$ and $\phi_{i}-\phi _{b}=5$


 We see that instead of the SD contour going off to infinity
 in the first quadrant of the first sheet, like in the previous model,
 it is now shifted towards $\xi =0$ by the singularity that exists
 there.

This is possible because as $\xi \rightarrow 0$ along the SD contour, the imaginary
 part of the action remains constant while the real part diverges to
 $+\infty $,

\be
 Re[I(\xi ,S,\phi _{i},\phi _{b})]\rightarrow
 \frac{S}{H^{2}}\biggl(\frac{N_{3}}{160\sqrt{3}}\biggr)(\phi _{i}-\phi _{b})^{2}\mid
 \xi \mid ^{-1}
\ee

Thus the contribution of the singularity to the SD integral is vanishing.

As we pass through the Lorentzian classical solution  $\overline{\xi
}_{I}\in (-\infty ,1/4]$ we emerge into the second sheet. From then on
the behaviour of the SD contour is very similar to the one in the pure
gravity model. There is a second branch cut crossing somewhere between
$1/4$ and $3/8$, (according to the value of $S$), and the SD contour
then proceeds to infinity along  the third  sheet. Again,
since
the asymptotic behaviour of the action  in the massless field models
and in the pure gravity models is the same,
 the localisation of the contours and the behaviour of  the real
part of the action  will be common to both models,
 and that guarantees the convergence of the SD integral in 
the massless scalar field models.

Obviously as in the previous model, the total SD contour also
includes   a similar SD contour (complex conjugate)  for the other classical
Lorentzian solution, located on the second sheet, ensuring that
the resulting wavefunction is real.

\subsection{Massive Scalar Field Model}

Although, as  in the massless case, we now have a different  SD contour for each
value of $S$, $\phi_{b}$ and $\phi_{i}$, an appropriate SD contour
leading to a Lorentzian classical spacetime can always be found, for
each value of $\phi_{i}$.    
We are mainly interested in what happens for large values of $S$, for
which there is a pair of Lorentzian classical solutions
$\{\overline{\xi}_{k}(S,\phi _{b}),\overline{\phi}_{i}^{k}(S,\phi
_{b})\}, k=I,II$, for every value of $S$ and $\phi _{b}$. Figures
$10$, $11$, and $12$
show the results of our numerical computation of 
the SD contour passing through the classical Lorentzian solution on
the first sheet,
$\overline{\xi}_{I}$, in the case of a cone over $\alpha _{4}$, ${\cal
S}^{2}\times {\cal S}^{1}$, and ${\cal T}^{3}$, for $S=100$ ,
$\phi_{b}=1$ and $m=1$.

They all have similar behaviours, and so we shall concentrate
on the  $\alpha _{4}$ case, in figure $10$. Proceeding upward from the extremum, $\overline{\xi}_{I}(S=100,\phi _{b}=1)=0.2114$
the SD contour proceeds to infinity along the first quadrant of  the plane, along
the parabola

\be
\frac{a_{3}+A(\phi _{i},\phi
_{b})}{H^{2}}S(S-S_{crit})Im(\sqrt{\xi})=\tilde{I}[\overline{\xi}_{I},\overline{\phi
_{i}}] 
\ee

The convergence of the integral along this part of the contour is
guaranteed by the asymptotic behaviour of the action along the SD
contour on the first sheet

\be
Re [I^{I}(\xi ,S,\phi _{i},\phi _{b})]\sim \frac{a_{3}+A(\phi _{i},\phi _{b})}{H^{2}}S(S-S_{crit})\sqrt{\mid \xi \mid}
\ee

Moving downward from the extremum at $\overline{\xi}_{I}=0.2114$, we
immediately  cross the branch cut and hence emerge onto the 
second sheet. Once again due to the alteration of the sign of the
action one cannot proceed immediately to infinity. Instead the SD contour
crosses the branch cut once more at $\xi =0.2763 $ between $1/4$ and $1/3$, emerging 
onto the third sheet, where it finally proceeds to infinity in  the
first quadrant of this third sheet along

\be
\frac{a_{3}+A(\phi _{i},\phi
_{b})}{H^{2}}S(S+S^{III}_{crit})Im(\sqrt{\xi})=\tilde{I}[\overline{\xi}_{I},\overline{\phi_{i}}] 
\ee

 Once more convergence is a consequence of the asymptotic
behaviour

\be
Re [I^{III}(\xi ,S,\phi _{i},\phi _{b})]\sim \frac{a_{3}+A(\phi _{i},\phi _{b})}{H^{2}}S(S+S_{crit}^{III})\sqrt{\mid \xi \mid}
\ee

As before, the SD contour that passes through the other extremum
located in the equivalent position, $\overline{\xi}_{II}=0.2114$, but in the
second sheet, is the complex conjugate of the previous  contour,
and the full SD contour is then taken to be the union of these two
sections. Thus we see that the semiclassical approximation based on
these two Lorentzian classical solutions can be justified in all three
cases by the existence of these SD contours.

Note that although we only present the results for the most
significant topologies the same computations can be performed for any
other simplicial spacetime that is a cone over a closed, connected
combinatorial $3-$manifold. We only have to change the values of
$N_{1}$ and $N_{3}$ accordingly.

\sect{Conclusions}

Having emphasised the need to consider more general spacetimes, other
than the ones that are classical solutions of Einsteins' equations,
i.e., manifold-based spacetimes, we then proceeded to show that the
contributions for the wavefunction of the universe coming from some of
these non-manifold based spacetimes are still consistent with a
wavefunction predicting classical spacetime for the late universe, in
accordance with our everyday experience. In order to have a finite
representation of these spacetimes we appealed to the simplicial
representation of topological spaces and the Regge calculus
formulation in order to specify the simplicial metric on those
simplicial complexes. 

By considering only complexes of  cone-like type (combinatorial 
conifolds), a natural minisuperspace approximation arises, with only
one internal metric degree of freedom. To make it somewhat more 
realistic we also considered the presence of a massive scalar field.

 Proving the existence of convergent steepest descent contours 
of integration associated with  Lorentzian classical  solutions
for each case, we justified the validity of the semiclassical 
approximation for the minisuperspace wavefunctions associated with our
models. The resulting wavefunctions  do indeed exhibit an
oscillatory behaviour for the late universe, consistent with the
prediction of Lorentzian classical spacetime as we know it.

These results  add credibility to the generalization
of the concept of history in the sum over histories formulation 
of quantum gravity to include more general, non-manifold based
histories, in particular conifold based histories. Conifolds are
particularly
suitable for our objectives because although more general than
manifolds they are still regular enough for all the usual basic concepts in
GR to be easily extended to them. On the other hand their situation
concerning algorithmic decidability is more favourable than that of
manifolds because the set of closed connected $4-$conifolds is known
to be
algorithmically decidable,
 which is still an open problem for $4-$manifolds.

Finally, it should be noted that of the three conditions that
 any contour of integration should verify, the SD contours we have
 calculated only explicitly obey two of them, namely, convergence and 
leading to a  wavefunction predicting classical space-time when the universe is
 large. We have not concerned ourselves with what regards the
 satisfaction of the  constraints implementing
diffeomorphism invariance, the status  of which is still not fully understood 
in the simplicial framework \cite{ruth2}. However we expect 
the minisuperspace approximation
to be a good testing ground for further study.

\vspace{12pt}

{\bf Acknowledgements}

The authors thank Kristin Schleich for helpful discussions and advice.
 The work of C. Correia da Silva was supported by the Portuguese
 Ministry of Science and Technology under grant PRAXIS XXI BD-5905/95.
Other support has come from the UK Particle Physics and Astronomy
 Research Council.

\end{document}